# MANDARIN SINGING VOICE SYNTHESIS BASED ON HARMONIC PLUS NOISE MODEL AND SINGING EXPRESSION ANALYSIS

*Ju-Chiang Wang*

Institute of Information Science, Academia Sinica

Taipei, Taiwan

asriver@iis.sinica.edu.tw

*Hung-Yan Gu*

Dept. of Computer Science and Information Engineering, National Taiwan University of Science and Technology

Taipei, Taiwan

guhy@mail.ntust.edu.tw

*Hsin-Min Wang*

Institute of Information Science, Academia Sinica

Taipei, Taiwan

whm@iis.sinica.edu.tw

**ABSTRACT**

The purpose of this study is to investigate how humans interpret musical scores expressively, and then design machines that sing like humans. We consider six factors that have a strong influence on the expression of human singing. The factors are related to the acoustic, phonetic, and musical features of a real singing signal. Given real singing voices recorded following the MIDI scores and lyrics, our analysis module can extract the expression parameters from the real singing signals semi-automatically. The expression parameters are used to control the singing voice synthesis (SVS) system for Mandarin Chinese, which is based on the harmonic plus noise model (HNM). The results of perceptual experiments show that integrating the expression factors into the SVS system yields a notable improvement in perceptual naturalness, clearness, and expressiveness. By one-to-one mapping of the real singing signal and expression controls to the synthesizer, our SVS system can simulate the interpretation of a real singer with the timbre of a speaker.

## 1. INTRODUCTION

To make a synthetic singing voice considered a natural, human-like sound is a challenging task. One important issue is how to impersonate a real human singing a song according to the lyrics and the associated score. As a human singer can interpret a song in his or her own way, a singing voice synthesis (SVS) system should be able to produce different interpretations of the same piece of music [1]. We believe that these interpretations result from a set of singing expressions; hence the relationship between the expressions and some control features is crucial to synthesizing an expressive singing voice. The purpose of this research is to analyze the parameters that control the expression of singing, so that we can simulate the singing voice of a popular singer with a generally agreed expression. The input singing voice signal is sung by a real singer who follows the main melody of the MIDI score, and the recording is then analyzed with the MIDI file of the popular song. The singer can perform the singing expression by following the interpretation of the original singer or interpret the expression in his or her own way. By collecting multiple versions of the same song, we can obtain different singing expressions that follow the same MIDI score. With the collected data, how a singer interprets a MIDI score with his or her expressive techniques can be analyzed, and the results can be used to design machines that sing the MIDI score expressively like humans. To fulfill this task, an in-depth analysis of human singing is necessary, and an SVS system is required. In this paper, we propose a framework for Mandarin expressive SVS, and describe an implementation of the SVS system.

We have been working on SVS for several years. In 2005, a Mandarin SVS system was developed by modeling the voiced-part of a syllable as additive sinusoids [2]. Specifically, the system synthesizes a singing voice signal by using the fundamental frequency (according to the MIDI note) and sinusoidal parameters. However, due to the lack of high-band noise in the voiced-part, the synthesized signal sounds artificial like that of vocoders [3]. Subsequently, the harmonic plus noise model (HNM) was adopted [4, 5]. HNM overcomes the above disadvantage and substantially improves the Mandarin SVS system in terms of perceptual naturalness and clearness. However, since it does not process singing expressions and emotions, the synthesized singing still sounds unrealistic and dull.

There are some related works by other research groups. Meron and Hirose [6] implemented a mechanism of vibrato singing on a large database containing units with vibrato for synthesis. By ensuring that the phase of the vibrato used for synthesis was consistent with the vibrato phase of the synthesized (target) vibrato sound, the required modification of the original unit's prosody could be minimized. Bonada *et al*. defined a set of musical controls to represent the expression of singing voices in their SVS system for Spanish [7, 8]. A commercial SVS system, called VOCALOID [9], developed by the YAMAHA Corp. was launched in 2004. Its score editor provides an integrated environment for users to input notes, lyrics, and expressions. In 2006, Janer *et al*. researched expression controls of SVS [10]. They used an analysis module to extract expressive information from the input singing voice signal, after which they adapted and mapped the internal synthesizer controls with the extracted information. Their goal was to develop a real-time performance-driven SVS system. Meanwhile, a corpus-based SVS system for Mandarin Chinese was proposed in [11, 12]. The authors designed three corpora for SVS and defined two distance functions. They applied the Viterbi search algorithm to identify optimal combinations of synthesis units from the three corpora, and combined the synthesized output with several sound effects.

The remainder of this paper is organized as follows. Section 2 introduces the factors that influence singing expression.

Section 3 contains an analysis of these expression parameters. In Section 4, we describe the proposed Mandarin expressive SVS system. The results of perceptual experiments are discussed in Section 5, and Section 6 summarizes our conclusions.

## 2. FACTORS THAT INFLUENCE SINGING EXPRESSION

The expression of singing may be influenced by many factors ranging from the structure of the song to each note in it. In this research, we focus on the factors at the musical note level in order to capture the acoustic features of the singing voice. Other factors such as musical structures (e.g., verse, chorus, bridge, etc) and musical marks (e.g., crescendo, diminuendo, accelerando, animato, etc) created by composers are not considered.

Mandarin Chinese is a syllable-timed language. There are only 408 distinct syllables if the tones are ignored. When singing a Mandarin song, a syllable from the lyrics may relate to a single musical note, or more notes with a portamento. In contrast, by definition, a note in a song can only correspond to one syllable, and notes that correspond to multiple syllables are separated into multiple notes. Therefore, in this study, syllables are used as the basic musical units (denoted as musical syllables).

Based on previous studies of singing expression [7, 9, 10] [2, 5, 13] and our own survey of the relationship between the singing in a wave format and the corresponding MIDI scores, we consider that six factors strongly influence the performance of singing expression. The factors are: (1) pitch curve, (2) allocation of within-syllable phonemes, (3) dynamics, (4) onset time, (5) features of sliding in a long musical syllable, and (6) timbre. They are modeled and implemented in our Mandarin SVS system.

*Pitch-curve:* The pitch curve of a musical syllable plays an important role in the expression of singing voices from the perceptual aspect, such as vocal sliding (portamento) or vibrato. Here, potamento is also applied like the "slide" or "bend" functions of instrument synthesizers. Prame [14] and Arroabarren [15] summarized the vibrato parameters, namely frequency, extent, and intonation, for Western songs. It is worth mentioning that pitch curves show up not only in syllables related to multiple notes, but also in one-note-related syllables because they may be influenced by the pitch of their neighboring notes.

*Allocation of within-syllable Phonemes:* A Mandarin syllable has a three-element structure, $C_x$-V-$C_n$. The first element, $C_x$, can be a voiced initial (consonant), an unvoiced initial (consonant), a glide (/i/ of /iau/), or null (nothing). The second element, V, is a vowel. It can be a monophthong (e.g., /i/), a diphthong (e.g., /ia/), or a triphthong (e.g., /iau/). The last element, $C_n$, can be either a nasal ending or null. When a Mandarin syllable is sung, it becomes duration-varying, and the changes in duration are different in each element. From our observations, the vowel element is usually the most varied part when the duration of a syllable changes. Hence, we further divide a vowel into three segments, namely, A(attack), S(sustain), and R(Release), following the concept widely used in computer music [16]. The A-S-R segmentation is shown in Fig. 1. Since a vowel may contain multiple phonemes, modifying only the duration of the S segment can ensure that the most important phoneme in a vowel will be perceptually intact. In our Mandarin SVS system, the duration-manipulation of a musical syllable is achieved by modifying the $C_x$, V, and $C_n$ elements with different ratios according to the phonemes and the principle described above.

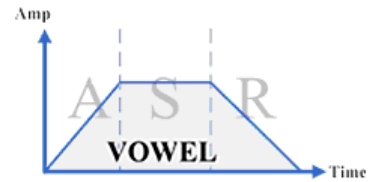

Figure 1: *Illustration of A-S-R segmentation in a vowel.*

*Dynamics:* The dynamics can be divided into two categories: the voiced-part dynamics and the unvoiced-part dynamics. The voiced-part dynamics is defined as the amplitude tendency of the voiced-part in a musical syllable, which is generally influenced by the syllable structure. It represents the amplitude of the voiced-part, which is more perceptually sensitive than the unvoiced-part. A vibrato usually occurs with a tremolo, which can be seen in the envelop. In future research, it would be interesting to investigate the relationship between vibrato and tremolo. The unvoiced-part dynamics is used to describe the loudness of the unvoiced initial (consonant) of a musical syllable.

*Onset Time:* Onset time refers to the temporal synchronization between a musical syllable and the related note(s). The onset time of a note is approximately synchronized at the stressed vowel element of a musical syllable. In consequence, both voiced and unvoiced initial (consonant) of a musical syllable usually appears earlier than the associated onset time [2, 5]. In a real performance, a time-shift between the stressed vowel and the onset time within a certain range is not considered out of beat, but shows the time-dynamic characteristics of the singer's grooves.

*Features of sliding in a long musical syllable:* Some prosodic variations may occur within a long musical syllable, especially when the long musical syllable is in the final position of a singing phrase. These variations, which sound like portamento, can be described by two features. One is the deviation between the singing pitch curve and the key of the corresponding MIDI score. Musical scores may not instruct the singer to bend the pitch; however, it happens naturally just like a smooth pitch-transition among the musical syllable's neighbors. The other feature is the repetition of the stressed vowel. The long musical syllable with vocal sliding may sound like a tight-concatenation of two vowels, where the second one is the stressed vowel of the first one. For example, when a long musical syllable /diau/ with vocal sliding is sung, the voice may sound like /diau-au/, where /au/ is on the bending segment.

*Timbre:* We define the timbre as the tone-variation between the samples in the SVS corpus and the synthesized signal. The corpus used for synthesis is recorded as the speech voice of each phoneme in Mandarin Chinese, rather than the singing voice. Hence, extra adjustments are made in the synthesis processing stage to simulate the timbre of the singing voice. The adjustments include strong or weak sound, brightness, and clearness of sounds [17, 9].

## 3. EXPRESSION PARAMETER ANALYSIS

In this section, we present our method for analyzing the singing expression. For the analysis, we need to record some real singing voices. The singing voice signal is then analyzed together with the MIDI score and the lyrics to reveal the relation between the signal and the expression represented by the MIDI-lyrics pair. By combining the results, an analysis module can produce a parame-ter set for synthesizing the singing expression. Fig. 2 illustrates the block diagram of the proposed expressive SVS framework. We discuss the collection of singing data and the analysis module in the following subsections, and then describe the singing voice synthesizer in Section 4.

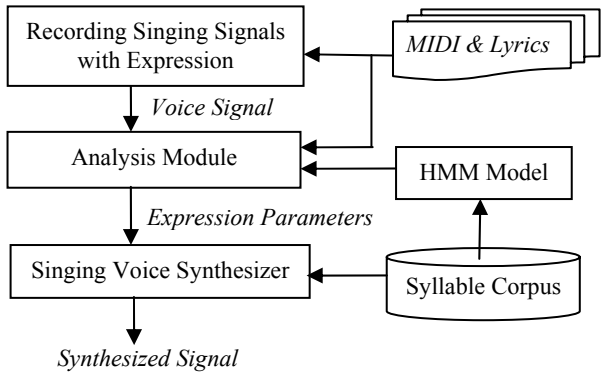

Figure 2: *The proposed expressive SVS framework.*

### 3.1. Recording of Human Singing Signals

We need the singing voices of popular songs for the analysis. As it is difficult to obtain the clean singing voice of a professional singer, we use the singing voice of an amateur singer instead. We invited a female (denoted as female_A) to record her singing voice signal in the format of 16 bits, 22,050 Hz, and mono. The recording was made in a professional soundproof room. We chose a popular sound module as the MIDI device. The MIDI channels of the background music and the main melody were separated. The channels of the background music were set as default virtual instruments. Note that the background music is important for the singer to know the tempo and key range of the current and upcoming sections. The channel of the main melody was set as a unique instrument synthesizer (with good sustain, stable pitch-contour and envelop) at a louder sound level for the singer to monitor. We made several recordings of each song and chose one with an explicit pitch and tempo as the main melody for analysis.

### 3.2. Segmentation of musical syllables

The segmentation of musical syllables is based on the time information of MIDI scores. We use the timing of note-on and note-off as the reference for musical syllable boundaries in the singing signal. Segmentation is performed by automatic HMM-based forced-alignment [18] and additional manual checking based on some acoustic features of frames around the timing of the MIDI score. This technique is based on the assumption that the singer followed the MIDI score exactly. The details of phoneme segmentation within musical syllables are presented in Section 3.5.

As mentioned earlier, a musical syllable may relate to a single MIDI note, or more notes with a vocal sliding. A musical syllable may be sung naturally with vocal sliding, but without an explicit instruction of portamento in the MIDI scores. This situation is commonly observed in the pitch curve. A more common situation is that MIDI scores instruct a singer to sing with portamento. When the succeeding note's note-on appears before the current note's note-off, it is defined as a sign of portamento.

Notes with this sign are combined as one segment to represent a musical syllable with portamento. We use the example in Fig. 3 to explain this principle. In our segmentation module, all MIDI files are normalized following this principle.

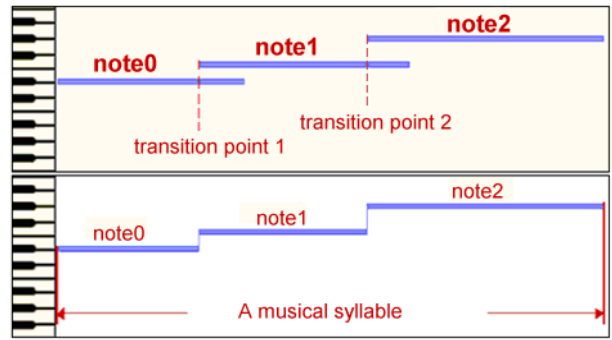

Figure 3: *note0, note1, and note2 are combined as one musical note for the segmentation of a vocal sliding syllable. Key and time information of each note are recorded.*

### 3.3. Pitch Curve Detection

Pitch curve can be represented by a sequence of fundamental frequencies extracted from a sliding analysis frame with size of 20 ms. The extraction of fundamental frequency $P$ is based on the combination of an auto-correlation function $R(k)$ and an absolute magnitude difference function (AMDF) $M(k)$ [19] given by

$$P = \arg\max_{P_{\min} < k < P_{\max}} \frac{R(k)}{M(k)+1}. \quad (1)$$

The range of pitch is from 60Hz to 500Hz. For any $k$ within the range, we pick the maximum $P$ as the estimated pitch of the frame. The key of the corresponding MIDI note helps to correct gross pitch errors (e.g., halving and doubling of pitch).

We employ two simple criteria to determine whether the frame is unvoiced: (a) $R(P)$ is smaller than one fourth of the energy (i.e., $R(0)$). (b) The quotient of $\max_{P_{\min} < k < P_{\max}}(M(k))$ and $\min_{P_{\min} < k < P_{\max}}(M(k))$ is smaller than 2.

### 3.4. Detection of Dynamics

The voiced-part dynamics can be considered as a curve representing the energy as a function of time. In practice, it is recorded as a sequence of frame energy. The unvoiced-part dynamics is represented as the maximum amplitude of the unvoiced-consonant segment, as mentioned in Section 2.

### 3.5. Phoneme Alignment and sub-segmentation in musical syllables

After identifying the boundaries of each musical syllable, we apply a HMM/SVM-based method [20, 18] for automatic phoneme segmentation. Then, A-S-R segmentation of V is based on the envelop curve of the vowel, which is represented by a sequence of the maximum amplitude $A_i$ of the $i$-th frame given by

$$A_i = \arg\max_{x[t] \in i\text{-th frame}} |x[t]|. \quad (2)$$

The segments of attack and release can be labeled by an adaptive threshold.

To improve the performance of the SVS system, we need explicit segmentation information; hence a further adjustment is required. The pronunciation of the initial ($C_x$) and the final ($V$-$C_n$) in Mandarin Chinese syllables can be classified into five categories and two categories, respectively. The five classes of the initial $C_x$ are: (I) stop, including /b, p, d, t, g, k/; (II) fricative, including /c, f, h, j, q, s, z/; (III) nasal, including /m, n/; (IV) glide, including /l, r, w, y/; and (V) null. In (I), the segment of stop is tuned with an obvious pulse in the spectral flux. In (II), the fricative is recognized by a gentle pulse (a pulse of a longer duration) of the zero-crossing rate. In (III), the boundary between the nasal initial and the vowel is tuned with the valley of the spectral variance. In (IV), the boundary between the glide and the vowel is the most ambiguous; therefore, it is tuned manually. The two categories of the final $V$-$C_n$ are: vowels with nasal ending and vowels without nasal ending. The nasal ending is labeled by automatic phoneme segmentation. A conceptual example of the syllable /man/ is illustrated in Fig. 4. Five segments in /man/ are labeled by the analysis module, namely, the voiced initial /m/, the attack, sustain, and release of the vowel /a/, and the nasal ending /n/. Fig. 5 shows an example of musical syllable segmentation, phoneme alignment, and the timing of the corresponding MIDI score. Musical syllables with a corresponding portamento instruction of MIDI can be manually assigned as multiple syllables, as discussed in "*Features of sliding in a long musical syllable*" in Section 2. Then, each syllable will have its own segmentation.

The timing of the stressed vowel, $t_v$, is defined as the beginning boundary of the V (vowel) segment in a musical syllable, e.g., the boundary between the initial consonant /m/ and the vowel /a/ in Fig. 4. The onset time of the corresponding MIDI note is denoted as $t_m$. Therefore, the time-shift discussed in Section 2 is represented by the deviation of $t_v$ and $t_m$ and recorded in the expression parameters.

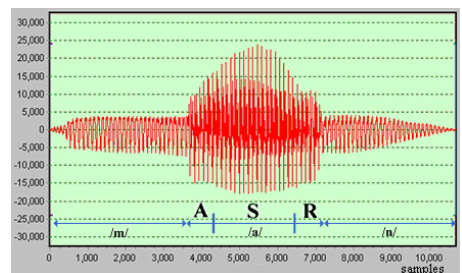

Figure 4: *Illustration of the five segments in the syllable /man/.*

vited to record 3,672 tokens of Mandarin syllables. Each syllable token was recorded by embedding it in the middle of a tri-syllable phrase (/a, i, u/ - syllable - /a, i, u/, $3 \times 408 \times 3 = 3,672$ combinations in total), and then chopped off semi-automatically. All syllables were uttered with the first tone. Pitch deviation was avoided during recording, and each syllable was uttered with the pitch contour as flat as possible. These samples were recorded with the same configuration as the singing signals described in Section 3.1.

Each syllable token was semi-automatically segmented following the principles discussed in Section 3.5. Then, for each syllable, the best token among the 9 samples was selected and the associated HNM parameters were extracted. Finally, the SVS corpus was comprised of these 408 labeled syllables together with their associated HNM parameters. The expression parameters extracted from the singing voice signal of the female_A singer were used to control the SVS system.

### 4.1. HNM

For a voiced speech frame, we can easily observe periodic characteristics from the spectrum: peaks occur at frequencies spaced by the fundamental frequency. However, these peaks only locate in a limited frequency range. In Fig. 6, the top spectrum is derived from the frame of the speech signal below it. When the frequency is less than 5,000Hz, the peaks (shown as little boxes) appear regularly. In contrast, upper 5,000Hz, the peaks appear unstably.

HNM [21] assumes that a speech signal is composed of a harmonic part and a noise part. The harmonic part accounts for the quasi-periodic component of the speech signal, while the noise part accounts for the non-periodic component. These two components are separated in the frequency domain by a time-varying parameter called the maximum voiced frequency (MVF).

## 4. SINGING VOICE SYNTHESIS

We selected the harmonic plus noise model (HNM) because of the high accuracy and flexibility of its frequency domain representation. The model has shown good results with timbre modifications of speech [21, 22] and singing voice signals [23]. Moreover, it has been adapted for Mandarin speech [22] and singing voice synthesis [4, 5]. A female (denoted as female_B) was in-

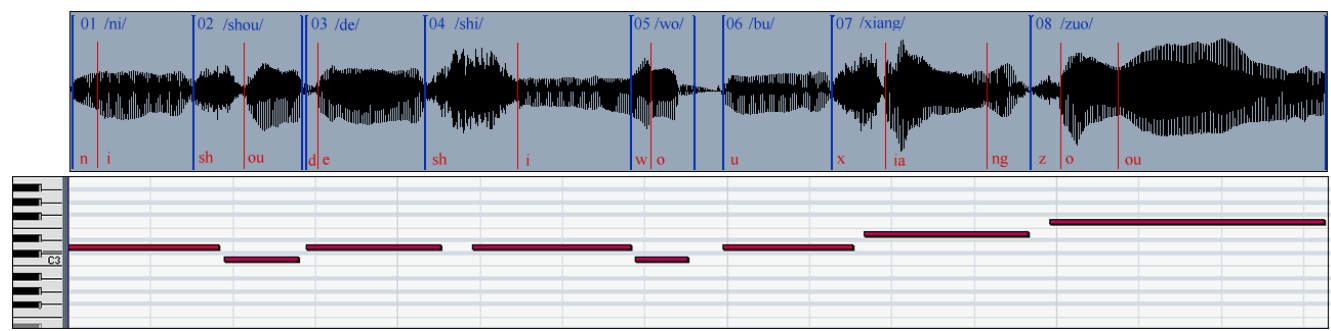

Figure 5: *An example of the musical syllable segmentation (long boundaries), phoneme alignment (short boundaries), and the timing of the corresponding MIDI note (the panel below). The singing signal sung by Female_A was from the first singing phrase of "Bad Boy – A-Mei Chang".*

The lower band of the spectrum (below MVF) is represented solely by the harmonics $h(t)$, while the upper band (above MVF) is represented by a modulated noise $n(t)$. Therefore, the analyzed signal $s(t)$ is expressed as:

$$s(t) = h(t) + n(t). \qquad (3)$$

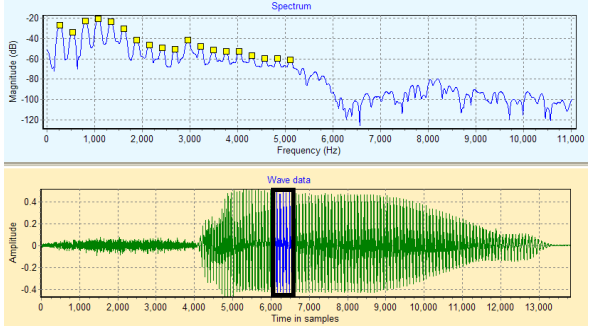

Figure 6: *A frame and its spectrum of a syllable /sha/.*

The lower band, or the harmonic part, is modeled as the sum of harmonics,

$$h(t) = \sum_{k=1}^{K(t)} a_k(t)\cos(\phi_k(t)), \qquad (4)$$

where $K(t)$ denotes the number of harmonics included in the harmonic part at $t$; $\phi_k(t)$ denotes the fundamental frequency; and $\alpha_k(t)$ denotes the amplitude of the $k$-th harmonic. For an unvoiced frame, MVF is set at zero; therefore, the whole band of the spectrum is viewed as a noise part. HNM assumes that the upper band of a voiced speech spectrum is dominated by modulated noise that can be modeled by harmonics with a constant fundamental frequency. For the implementation, we estimate the amplitudes of the harmonics with frequencies 100Hz apart. In other words, we use 100Hz as the fundamental frequency for estimating the HNM parameters. Then, these amplitudes are transformed into cesptrum coefficients to represent the smoothed spectrum of the noise part.

In the synthesis stage, with accurate HNM parameters, such as the fundamental frequency, MVF, the amplitude and phase information of each harmonic, and the cesptrum coefficients, the synthetic signal is constructed as:

$$\hat{s}(t) = \hat{h}(t) + \hat{n}(t). \qquad (5)$$

### 4.2. Time Mapping of Segments

Before synthesizing the signal, we need to set the boundaries of each segment in a synthetic syllable. We use linear time mapping of segments to locate the correct timing and duration of phonemes in a syllable. This technique ensures that the phoneme timings within a synthetic syllable match that within the analyzed syllable involved in the expression parameters. To implement this task, each syllable sample in the SVS corpus and the singing voice signal must be labeled by the process described in Section 3.5. In Fig. 7, **X** is a source syllable sample in the SVS corpus, where x1 and x5 denote the consonant-initial and nasal-ending segments, respectively; and x2, x3, and x4 denote the attack, sustain, and release segments, respectively. The corresponding segments from a target musical syllable **Y** are denoted as y1~y5, respectively. Therefore, the relative pairs $(x_i,y_i)$, $i=1,…,5$, represent the linear mapping relations on the time axis. In this case, the mapping is one-to-one; therefore, the boundaries of the synthetic signal are set individually according to the boundaries associated with y1~y5 recorded in the expression parameters.

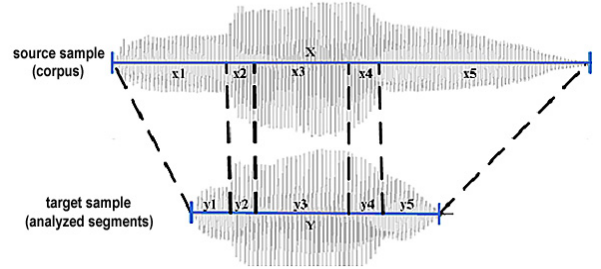

Figure 7: *Illustration of linear time mapping from the source sample to the target sample, based on the analysis of the musical syllable.*

### 4.3. Control Points and HNM Parameters

After the time mapping of segments has been completed, "Control Points" are set on the target time axis to extract the HNM parameters from the source syllable sample in the SVS corpus [22]. Thus, each control point has its own mapping instance in the source sample. We allocate control points with equal time intervals (every 100 samples, i.e., 4.54ms) on the synthetic time scale. The HNM parameters of a control point are determined by linear interpolation of the HNM parameters of the mapped source frames. This method is based on the continuity of the HNM parameters between two analyzed frames. The HNM parameters of the remaining target signal samples (i.e., the other 99 samples) can be estimated by linear interpolation of the HNM parameters of each sample's preceding and succeeding control points. This approach ensures that the synthesized signal will be continuous (smoothed), and also reduces the computational overhead.

### 4.4. Pitch Curve Control

In addition to the HNM parameters, the pitch curve, which is one of the expression parameters, must also be aligned at the control points. Cubic Spline interpolation is employed for the alignment of the analyzed and synthesized time scales. To maintain the original timbre during this adjustment operation, it is necessary to reconstruct envelops of the original spectrum and phase from the HNM parameters. Otherwise, if we only change the frequency of each harmonic in HNM, the wave envelop will not be maintained and the synthesizer might output a child-like voice, as shown in Fig. 8.

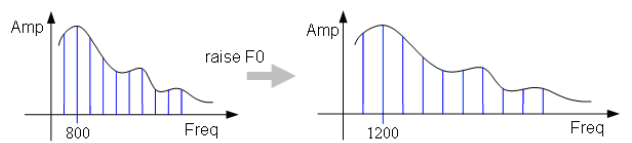

Figure 8: *Raising the frequency of each harmonic without estimating new amplitudes will change the spectrum envelop.*

For the implementation, we estimate the target amplitude and phase by interpolating the source amplitudes and phases of each harmonic using Cubic Spline. We use phase unwrapping to avoid interpolation errors, which would result in phase discontinuity. In musical syllables, pitch-shift is given by

$$\hat{C} = C \times 2^{\frac{k}{12}}, \quad (6)$$

where $\hat{C}$ is the shifted pitch curve of the original curve C, and $k$ is an integer that denotes the rise or fall of the key in semitones. A positive $k$ represents a rise, and a negative k represents a fall.

### 4.5. Co-articulation Simulation

If there is no short pause (i.e., voice onset time (VOT), which often happens when the succeeding syllable has a stop initial) between two concatenated musical syllables, we use two schema to smooth the discontinuity and to simulate the transition of two tight-concatenated musical syllables. First, a voiced-transition occurs when a syllable is followed by another syllable with a voiced initial. In this case, the final element of the preceding syllable and the initial element of the succeeding syllable are extended proportionally for cross-fading, as shown in Fig. 9. Second, in the case that a syllable is followed by another syllable with a fricative initial, the fricative initial of the succeeding syllable is extended to overlap a small part of the preceding syllable's final. These extended parts are processed together with their originating syllables, as described in Section 4.2, although they were generated in previous planning stage.

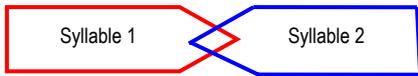

Figure 9: *Cross-fading of the voiced-transition.*

### 4.6. Dynamics Control

After the above five preparatory steps (i.e., segment alignment, control points allocation, estimation of the HNM parameters, pitch curve controls, and cross-fading), a musical syllable is synthesized in the format of 16 bits, 22,050Hz and mono. We then apply some post-processing steps for dynamics control. First, the dynamics of a synthesized musical syllable is normalized. The amplitudes of the voiced-part are adjusted to the level of the target (analyzed) musical syllable. In addition, the extended parts are adjusted linearly in fade in/fade out fashion. For the unvoiced-part, the signal is formed by noise, and the energy curve control has little effect on perception. Therefore, we linearly adjust the amplitudes of the unvoiced-part in proportion to the maximum amplitude of the target (analyzed) musical syllable recorded in expression parameters.

### 4.7. Concatenating Synthesized Musical Syllables

Finally, the synthesized musical syllables are concatenated to form a singing phrase. The concatenation is based on the characteristics and onset time of each synthesized musical syllable, as shown in Fig. 10. In the upper panel of the figure, P1~P4 denote the onset times of sylb1~sylb4, respectively; and the synthesized signal of the singing phrase is shown in the lower panel. There is a voiced-transition between sylb2 and sylb3, so an overlap can be observed. Since slyb4 is a fricative-initial syllable, it is allo- cated before its onset time P4, and extended to touch sylb3's final element. There is a vocal sliding final element, sylb5, which is a repetition of the stressed vowel of sylb4 in the singing phrase. The concatenation of sylb4 and sylb5 can be viewed as a voiced-transition.

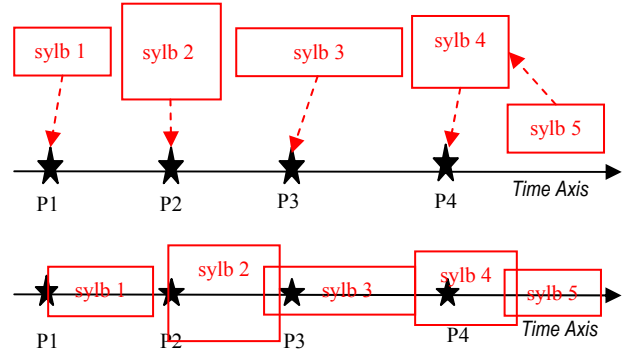

Figure 10: *Illustration of concatenating synthesized musical syllables.*

### 5. EXPERIMENTS AND DISCUSSIONS

We collected several singing voices in the format of 16 bits, 22,050Hz and mono using the same MIDI score and lyrics. The expression factor "timbre" is not considered in the experiments, which means we only used the original timbre of the SVS corpus. We evaluate three types of singing voices in the experiments, as shown in Table 1. Type (I) is the original singing voice of the real singer female_A. Type (II) is a synthesized signal obtained by considering the MIDI score, the lyrics, and the expression parameters extracted from type (I) simultaneously. Type (III) is a synthesized signal without any information of expression parameters, and it is based on the MIDI score and lyrics, linear manipulation of the duration of the segments in a syllable, and linear interpolation of the pitch-transition of vocal sliding. The three types of singing voices are shown in Fig. 11.

Table 1: *Type of singing voices in the experiments.*

| Type | I | II | III |
|---|---|---|---|
| Expression Parameters | Real singing voice | Yes | No |

The scoring of the perceptual experiment was as follows: the real singing voice was given a score of 5, and listeners were asked to score the other two samples in a range from 0 to 5 based on naturalness, clearness, and expressiveness. The real singing voice was presented to the evaluators first, and then the two synthesized singing voices were presented in random order, i.e., the evaluators did not know the label, (type II) or (type III), of the singing voice they heard. We synthesized four singing clips, each of which contained more than 7 phrases (i.e. more than 40 syllables) from Mandarin popular songs. These songs included "至少還有你" by Sandy Yi-Lam Lin, "Bad Boy" and "姊妹" by A-Mei Chang, and "執迷不悔" by Fei Wang. Two of the songs were fast, and the others were slow. Thirty adults not familiar with SVS and without any known hearing problems were invited to participate in the evaluation. The experiment results are shown in Table 2.

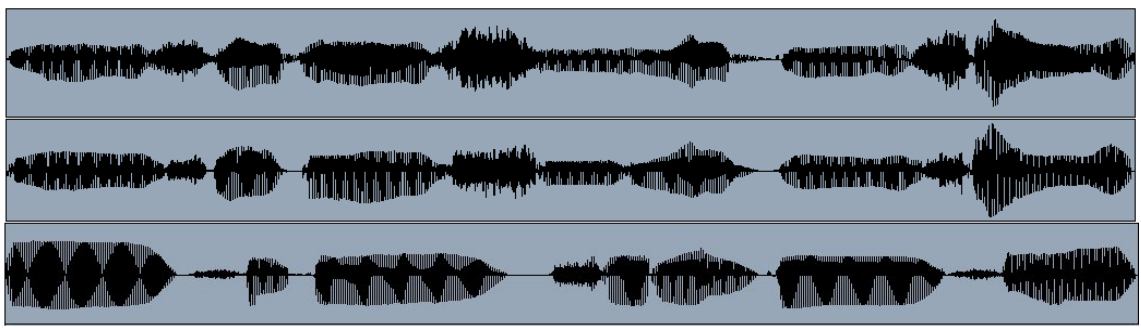

Figure 11: *Three different types of singing voice signals from the song "Bad Boy": from the top, the real singing (Type I), the synthesized singing signal generated with the expression parameters from the real singing (Type II), and the synthesized generated without using the expression parameters (Type III). All the three singing signals are obtained based on the same MIDI score and lyrics.*

Table 2: *The results of perceptual experiments.*

| Song Name | Type II | Type III |
|---|---|---|
| 至少還有你(slow) | 3.53 | 2.57 |
| Bad Boy (fast) | 3.27 | 2.27 |
| 姐妹 (slow) | 3.63 | 2.5 |
| 執迷不悔 (fast) | 3.47 | 2.53 |

The results demonstrate that the expression parameters substantially improve the SVS system's performance. Moreover, the results show that slow songs are preferable to fast songs. This may be due to two factors: (1) the lack of sample units; and (2) the lack of phase synchronization and dynamics smoothing in the cross-fading segments during the voiced transition of two syllables. We cannot minimize the spectral distance between two concatenated syllables because each syllable has only one available unit in the SVS corpus. Signals around the boundaries of a syllable are usually unstable; therefore, simply applying cross-fading may cause some clicks and noise during the voiced-part transition. These synthesized singing voices, including two complete songs and four clips, are available at "http://sovideo.iis.sinica.edu.tw/MSVS/".

## 6. CONCLUSIONS

In order to design machines that can sing like humans, we tried to investigate how human beings interpret musical scores and lyrics expressively. We attempted to represent the interpretations of a real singer to a specific song as the expression parameters to control the synthesizer. The expression parameters can be viewed as a set of low-level controls resulting from certain interpretations at an abstract level. We derived six factors related to singing expression that could lead to generally agreed interpretations. These factors were proposed to build the analysis module of a Mandarin Chinese SVS system. The results of perceptual experiments show that integrating the expression factors into the SVS system yields a notable improvement in perceptual naturalness, clearness, and expressiveness. By one-to-one mapping of the real singing signal and expression controls to the synthesizer, our SVS system can simulate the interpretation of a real singer (female_A) with the timbre of a speaker (female_B).

In our future work, we will exploit more useful expression factors and employ them in our SVS system. In addition, we will add a unit selection module to improve the fluency of concatenation. We also plan to build a large SVS corpus, and design an efficient cost function.

Our ultimate goal is to build an interpretation model of a classic singer. Even though the classic singer may no longer exist, we can design a virtual singer from the expression information of his/her left singing recordings. The virtual singer should be able to sing a new song with its own specific interpretation (i.e., specified parameters in the built model). However, there are still several challenging problems to be overcome. First, there is not yet a standard Mandarin singing corpus of the singer, or a sufficiently large collection of clean singing voices. Second, the segmentation accuracy of our analysis module needs to be improved. Third, an appropriate model for a large amount of expression parameters needs to be designed.

## 7. ACKNOWLEDGMENTS

This work was supported by Taiwan e-Learning and Digital Archives Program (TeLDAP) sponsored by the National Science Council of Taiwan under Grant: NSC 96-3113-H-001-012. The authors would like to thank Ting-Shuo Yo for fruitful discussions of writing and presentation, Huang-Liang Liao and Yen-Zuo Zhou for developing the HNM synthesizer, Zhi-Wei Zhou for recording the SVS corpus, and Rou-Lan Yan and Zheng-Yu Lin for recording the singing voices.